\documentstyle[prl,aps,epsbox]{revtex} 
\begin{document} 
\draft 
\preprint{} 
\twocolumn[\hsize\textwidth\columnwidth\hsize\csname@twocolumnfalse%
\endcsname 
\title{ 
Equivalence of Local and Separable Realizations of 
the Discontinuity-Inducing Contact Interaction 
and Its Perturbative Renormalizability 
} 
\author{ 
Taksu Cheon${,}^{1}$ 
T. Shigehara${ }^{2}$ 
and 
K. Takayanagi${ }^{3}$ 
} 
\address{ 
${ }^1$ 
Laboratory of Physics, Kochi University of Technology, 
Tosa Yamada, Kochi 782-8502, Japan\\ 
${ }^2$ 
Department of Information and Computer Sciences, 
Saitama University, 
Urawa, Saitama 338-8570, Japan\\ 
${ }^3$ 
Department of Physics, 
Sophia University, 
Chiyoda, Tokyo 102-8554, Japan\\ 
} 
\date{December 1, 1998} 
\maketitle 
\begin{abstract} 
We prove that the separable and local approximations of the 
discontinuity-inducing zero-range interaction in one-dimensional 
quantum mechanics are equivalent. 
We further show that the interaction allows the perturbative 
treatment through the coupling renormalization. 
%

\vspace*{3mm} 
\end{abstract} 
\vspace*{3mm} 
%
] 
%
%
%
\section{Introduction} 

The contact force different from the $\delta$-function 
interaction is an object 
that appears to have defied the intuition since the 
recognition of its existence more than 
a decade ago \cite{GK85,SE86,SE86a,AG88}. 
This second kind of contact interaction is an object found only in 
the quantum mechanics of one-dimensional particles.  It is 
characterized by the discontinuity of {\it wave functions themselves} 
at the location of the interaction which is contrasted to the 
discontinuity of the derivative of wave functions 
present in the ``usual'' $\delta$ function potentials. 
 From its start, this second kind of contact interaction 
has been conceived in abstract mathematical settings which are 
rather detached from any particular physical model. 
Because of the barrier of arcane mathematical language, 
it has been mostly hidden from the view of 
the physics community at large. 

There are some welcome signs 
which indicate a change in this situation. 
One is the recent discoveries of several non-trivial properties 
of the quantum systems with discontinuity-inducing 
interactions \cite{AE94,EX96,CH98,CS98a}. 
Also, there have been several attempts to represent
the interaction in terms of physically realizable
potential models \cite{CA93,CH93}.  Here, 
we single out one example in which
the discontinuity-inducing 
contact interaction is constructed in terms of elementary
local self-adjoint operator 
(``epsilon'' potential)  \cite{CS98}.
Its potential significance is in the fact that it can lead 
to the experimental manufacturing of wave function 
discontinuity and Neumann boundary. 
It appears that we are now ready for 
the application of the second kind of contact interaction 
to quantum system with practical relevance, in particular 
to the many-body systems \cite{CS98a}. 
One stumbling block to this end is the apparent non-perturbative 
nature of the interaction which can be recognized immediately by 
observing the divergence of the matrix element of the 
epsilon potential at the zero-size limit. 
None of the advanced field theoretical technique would be 
available without the representation of the interaction 
in the second-quantized form. 

Interestingly, we realize that 
the earliest physical realization of 
the discontinuity-inducing interaction by {\v S}eba 
as a separable potential 
with momentum dependent form factor 
(``prime-delta-prime'' function) \cite{SE86a} is of 
a great relevance, since it rests on the concept of 
renormalization with the use 
of a coupling constant which disappears at the zero-range 
limit. 
It is evident at this point, that the clarification of 
the relation between ``epsilon'' and ``prime-delta-prime'' 
representations is called for. 
Also, it is helpful to have a closer look at the workings 
of the discontinuity-inducing interaction 
in the context of the perturbative treatment. 

These two matters are exactly what we address in this paper. 
We first prove that the local and separable 
approximations of the 
discontinuity-inducing zero-range force in one-dimensional 
quantum mechanics are equivalent. 
We then look at the example of the spectra of 
a particle on a line with a contact force. 
One observes that the use of renormalized coupling results in 
the order-by-order cancellation of divergent term, and allows 
the perturbative treatment of the problem. 

It is worthwhile to recall the zero-range limit of small obstacle 
quantum mechanics in other dimensions than one.  As is well known, 
something resembling to $\delta$-function can be defined only in weaker 
sense with renormalized couplings \cite{AG88,SC97}. 
In hindsight, the fact that the simple $\delta$-function 
limit exists for small size obstacle in one-dimensional 
quantum mechanics is an accident, which could be either seen as lucky 
or unlucky.  The former view needs no explanation.  The latter view 
seems equally plausible since this has 
in effect delayed the wide recognition of 
the discontinuity-inducing interaction as an indispensable 
element of the one-dimensional quantum mechanics. 
The situation needs to be rectified. 
We would feel our purpose fulfilled 
if the current work serves as a step stone to 
that road. 

This paper is organized as follows.  In the next Section, 
we present an elementary rederivation of the result of 
{\v S}eba that the discontinuity-inducing interaction can 
be expressed as a momentum dependent separable form. 
In Section III, the equivalence of {\v S}eba's expression and 
the local expression is shown. Two coupling constants, bare 
and renormalized, are introduced. 
In Section IV, it is shown that the perturbative treatment 
of the discontinuity-inducing interaction brings about the 
divergence at the level of second order and beyond. 
This divergence, however, is shown to be made manageable 
with the coupling constant renormalization.

\bigskip 
%
\section{Rederivation of Seba's Prime-Delta-Prime Interaction} 
%

The discussion on discontinuity-inducing interaction 
traditionally started with the mathematical theory of 
self-adjoint extension. 
Here, we take another route which requires nothing fancier than 
the concept of $\delta$-function as the zero-size 
limit of regular potential with constant volume integral. 

We start by defining a function 
\begin{eqnarray} 
\label{2.1} 
\Delta_a(x) &=& {1 \over {2a}}  \ \ \ \ \ \ \ \ |x|\le a 
\\ \nonumber 
& &\ 0 \ \ \ \ \ \ \ \ \ |x|>a, 
\end{eqnarray} 
whose values at $x=\pm a$ are defined as the limiting 
values from $ \vert x \vert <0$ region. 
It has the property 
\begin{eqnarray} 
\label{2.2} 
\int_{-\infty}^{\infty}{dx \Delta_a(x)} = 1 . 
\end{eqnarray} 
Obviously, one has the Dirac's $\delta$-function as the zero-range 
limit; 
\begin{eqnarray} 
\label{2.3} 
\Delta_a(x) \rightarrow \delta(x) 
\ \ \ \ 
(a \rightarrow 0) . 
\end{eqnarray} 
 From the relation 
\begin{eqnarray} 
\label{2.4} 
\left( \Delta_a(x)\right) ^n 
= \left({1 \over {2a}}\right) ^{n-1} \Delta_a(x) , 
\end{eqnarray} 
one has 
\begin{eqnarray} 
\label{2.5} 
\Delta_a(x) f(\Delta_a(x)) 
= \Delta_a(x) f({1 \over {2a}}) . 
\end{eqnarray} 
We consider a wave function $\phi(x)$ that satisfies 
\begin{eqnarray} 
\label{2.6} 
-{{d^2}\over{dx^2}} \phi(x) + v \Delta_a(x) \phi (x) 
= k^2 \phi (x) . 
\end{eqnarray} 
At $a \to 0$, this wave function obviously 
satisfies the connection condition 
\begin{eqnarray} 
\label{2.7} 
\phi'(0_+) - \phi'(0_-) = v \phi(0_+) = v \phi(0_-) . 
\end{eqnarray} 
We now define another wave function $\psi(x)$ by 
\begin{eqnarray} 
\label{2.8} 
\psi(x) \equiv  {{d}\over{dx}} \phi(x) , 
\end{eqnarray} 
and look for the wave equation for $\psi(x)$, 
and study its properties. 
 From Eqs. (\ref{2.6}) and (\ref{2.8}), we have 
\begin{eqnarray} 
\label{2.9} 
\phi(x) = -\left( {1\over{k^2 - v\Delta_a(x)}} \right) 
{{d}\over{dx}}\psi(x) . 
\end{eqnarray} 
Note that the R.H.S. of Eq. (\ref{2.9}) is a product of two 
discontinuous quantities at $x= \pm a$, yielding the 
continuous quantity $\phi(x)$. 
Connection condition Eq. (\ref{2.7}) can be rewritten in terms of 
$\psi(x)$ as 
\begin{eqnarray} 
\label{2.10} 
\psi(0_+) - \psi(0_-) = -{v \over k^2} \psi'(0_+) 
                    = -{v \over k^2} \psi'(0_-) . 
\end{eqnarray} 
%
Differentiating Eq.(\ref{2.6}), one has 
\begin{eqnarray} 
\label{2.11} 
-{{d^3}\over{dx^3}} \phi(x) 
- v {{d}\over{dx}} \Delta_a(x) 
\left( {1\over{k^2 - v\Delta_a(x)}} \right) 
\phi (x) 
\\ \nonumber 
= k^2 {{d}\over{dx}}\phi (x) . 
\end{eqnarray} 
We obtain an equation for $\psi(x)$ in the form 
\begin{eqnarray} 
\label{2.12} 
-{{d^2}\over{dx^2}} \psi(x) 
-\left( {v\over{k^2 - v/(2a)}} \right) 
{{d}\over{dx}} \Delta_a(x) {{d}\over{dx}} \psi (x) 
\\ \nonumber 
= k^2 \psi (x) . 
\end{eqnarray} 
We define 
\begin{eqnarray} 
\label{2.13} 
c = -{v\over k^2} , 
\end{eqnarray} 
\begin{eqnarray} 
\label{2.14} 
c_a = -{v\over {k^2-v/(2a)}} . 
\end{eqnarray} 
 From these definitions, we have a relation 
\begin{eqnarray} 
\label{2.15} 
{1\over c_a} = {1\over c} + {1 \over {2a}} . 
\end{eqnarray} 
Rewriting Eq. (\ref{2.12}), 
we now have a Schr{\" o}dinger equation 
\begin{eqnarray} 
\label{2.16} 
-{{d^2}\over{dx^2}} \psi(x) + c_a E_a(x) \psi (x) 
= k^2 \psi (x) 
\end{eqnarray} 
with a potential 
\begin{eqnarray} 
\label{2.17} 
E_a(x) \equiv {{d}\over{dx}} \Delta_a(x) {{d}\over{dx}} . 
\end{eqnarray} 
One can see, from Eqs. (\ref{2.10}) 
and (\ref{2.13}), that 
the zero range limit $a \to 0$ of this potential 
results in the connection condition 
\begin{eqnarray} 
\label{2.18} 
\psi(0_+) - \psi(0_-) = c \psi'(0_+) = c \psi'(0_-) , 
\end{eqnarray} 
when the coupling $c_a$ is rescaled according to the relation, 
Eq.(\ref{2.15}). 

The operation of the $E_a$ potential can be interpreted in two ways: 
\begin{eqnarray} 
\label{2.19} 
\left\langle \psi_2 \right| E_a \left| \psi_1 \right\rangle 
=\left\langle \psi_2 \right| 
       {{d}\over{dx}}\Delta_a 
       {{d}\over{dx}}\psi_1 \left.\right\rangle 
=-\left\langle \psi_2'\right| 
  \Delta_a\left|\psi_1'\right\rangle . 
\end{eqnarray} 
The second equiation gives the expression 
\begin{eqnarray} 
\label{2.20} 
E_a(x) = -{ 
          {\mathord{\buildrel{\lower3pt\hbox 
            {$\scriptscriptstyle\bf\leftarrow$}}\over d} 
          } 
          \over{dx}} \Delta_a(x){{\vec d}\over{dx}} . 
\end{eqnarray} 
This is essentially the form with which 
the discontinuity-inducing contact 
interaction has been first realized physically 
by {\v S}eba \cite{SE86a}.

\bigskip 
%
\section{The Equivalence Between Prime-Delta-Prime 
and Epsilon Functions} 
%
In this Section, we prove that the separable expression, 
which we have rederived in the last section, is identical to 
the local realization developed in Ref. \cite{CS98}. 
We rewrite Eq. (\ref{2.11}) 
in terms of $\psi (x)$ as 
\begin{eqnarray} 
\label{3.1} 
-{{d^2}\over{dx^2}} \psi(x) 
+c_a 
\left({{d}\over{dx}} \Delta_a(x)\right) {{d}\over{dx}} \psi (x) 
\\ \nonumber 
+c_a 
\Delta_a(x) {{d^2}\over{dx^2}} \psi (x) 
= k^2 \psi (x) .
\end{eqnarray} 
The second term can be rewritten in two steps;  first 
\begin{eqnarray} 
\label{3.2} 
{{d}\over{dx}}\Delta_a(x) 
= {1\over {2a}} 
  (\delta(x+a)-\delta(x-a)) , 
\end{eqnarray} 
and then 
\begin{eqnarray} 
\label{3.3} 
& &{c_a\over{2a}} \delta(x+a){{d}\over{dx}}\psi(x) 
\\ \nonumber 
& & \ \ \ \ \ \ \ \ \ \ \ \ \ \ \ \ 
\approx -c_a\cdot{{\delta(x+a_+)-\delta(x+0_+)}\over {2 a^2}} \psi(x) 
\\ \nonumber 
& &{c_a\over{2a}} \delta(x-a){{d}\over{dx}}\psi(x) 
\\ \nonumber 
& & \ \ \ \ \ \ \ \ \ \ \ \ \ \ \ \ 
\approx c_a\cdot{{\delta(x-a_+)-\delta(x-0_+)}\over {2a^2}} \psi(x) . 
\end{eqnarray} 
Here, a care must be taken to handle the quantity 
$\psi'(x)$ which is discontinuous at $x=\pm a$. 
The above argument is based on the convention 
that the quantities at $x=\pm a$ is evaluated as the 
limiting values from $ \vert x \vert< a$ region, 
in accordance with Eq. (\ref{2.1}). 
If one adopts the alternative definition 
of the step function such that 
$\Delta_a(\pm a) = 0$, one has 
to replace $c_a$ in the second term of Eq. (\ref{3.1}), 
and thus in L.H.S. of Eq. (\ref{3.3}), 
by $c$ (see Eqs. (\ref{2.9}),(\ref{2.11})).
However, the relations $c\psi'(a_+)$ $= c_a\psi'(a_-)$ 
and $c\psi'(-a_+)$ $ = c_a\psi'(-a_-)$ 
guarantees that we still obtain the same result,
R.H.S. of  Eq. (\ref{3.3}). 
Operations become easy after this, since 
everything is expressed in terms of $\psi(x)$ which is 
continuous everywhere. 
We obtain 
\begin{eqnarray} 
\label{3.4} 
& &
-\{1-c_a\Delta_a(x)\} 
{{d^2}\over{dx^2}} \psi(x) 
\\ \nonumber 
& &
-{c_a\over{2a^2}} 
\{\delta(x+a_+)+\delta(x-a_+)-2\delta(x) \} \psi (x) 
\\ \nonumber  
&=& k^2 \psi (x) . 
\end{eqnarray} 
We divide this by $(1-c_a\Delta_a(x))$.  
Since this factor acts only within 
$|x|<a$,  we have 
\begin{eqnarray} 
\label{3.5} 
-{{d^2} \over{dx^2}} \psi(x) 
-{c_a\over{2a^2}} 
\{\delta(x+a)+\delta(x-a) \} \psi (x) 
\\ \nonumber 
+{c_a/a^2 \over{1-c_a\Delta_a(x)}} 
 \delta(x) \psi (x) 
\\ \nonumber 
= {k^2 \over{1-c_a\Delta_a(x)}} \psi (x) .
\end{eqnarray} 
This can be written in the form 
\begin{eqnarray} 
\label{3.6} 
-{{d^2}\over{dx^2}} \psi(x) 
+\varepsilon_a(x;c) \psi (x) 
= k^2 \psi (x) ,
\end{eqnarray} 
where 
\begin{eqnarray} 
\label{3.7} 
\varepsilon_a(x;c) 
= -{c_a\over{2a^2}} [ \delta(x+a)+\delta(x-a) ] 
\\ \nonumber 
+ {c_a\over{a^2}}\cdot {1\over{1-c_a/(2a)}}\delta(x) 
\\ \nonumber 
- {c_a\over{2a}}\cdot {k^2\over{1-c_a/(2a)}}\Delta_a(x) 
\end{eqnarray} 
represents the equivalent local potential of $E_a(x)$. 
Using the expressions 
\begin{eqnarray} 
\label{3.8} 
{c_a\over{2a}} &\approx& 1 - {{2a}\over c} + O(a^2), 
\\ \nonumber 
{c_a\over{1-c_a/(2a)}} &=& c , 
\end{eqnarray} 
one arrives at 
\begin{eqnarray} 
\label{3.9} 
\varepsilon_a(x;c) 
= \left( {2\over c}-{1\over a} \right) 
 \{ \delta(x+a)+\delta(x-a) \} 
+ {c\over{a^2}} \delta(x) . 
\end{eqnarray} 
This is exactly the $\varepsilon$-function, 
a local expression given in Ref. \cite {CS98} 
for the discontinuity-inducing interaction 
in one-dimensional quantum mechanics. 

\bigskip 
%
\section{Perturbative Renormalizability Up to Second Order} 
%
Despite its intuitive nature and usefulness in ``engineering'' 
purpose, the $\varepsilon$-function expression has 
a major setback in its apparent inability to 
cope with perturbative approach.  One realizes this fact easily 
by calculating the matrix element of $\varepsilon_a(x;c)$. 
For one thing, the coupling constant $c$ appears in its inverse. 
The limit $a \to 0$ immediately gives the divergence. 
The separable expression $c_a E_a(x)$ looks promising 
in that respect.  In this section, we show that the perturbation 
in terms of $c_a$ along with the renormalization procedure indeed 
gives the sensible answer through the explicit calculation of 
energy eigenvalue up to second order. 

Consider 
\begin{eqnarray} 
\label{4.1} 
-{{d^2}\over{dx^2}} \psi(x) + V(x) \psi (x) 
= k^2 \psi (x) 
\end{eqnarray} 
with $V(x)$ being either $v \Delta_a(x)$ 
or $c_a E_a(x)$ with disappearing $a$. 
For definiteness, we place the system on the line of length $L$ with 
periodic boundary 
\begin{eqnarray} 
\label{4.2} 
\psi(-L/2) = \psi(L/2), \ \ \ 
\psi'(-L/2) =  \psi'(L/2). 
\end{eqnarray} 
The problem now becomes that of free wave equation on a ``ring'' 
with a defect at $x = 0$ whose characteristics is specified by 
the zero-range function $V(x)$. 
The solution which satisfies the condition Eq. (\ref{4.2}) 
takes the form 
\begin{eqnarray} 
\label{4.3} 
\psi(x) &=& A \sin{k(x-\eta)} 
\ \ \ \ \ \ \ \ \ (x >0 ) 
\\ \nonumber 
        & & A \sin{k(x-\eta+L)} 
\ \ \ \ (x <0 ) , 
\end{eqnarray} 
where $A$ and $\eta$ are the constants which is to be determined. 
For $V = v \Delta_a(x)$ with $a \to 0$, 
the connection condition Eq. (\ref{2.7}) gives 
\begin{eqnarray} 
\label{4.4} 
\tan{ {{k L}\over 2} } = {v\over{2k}}. 
\end{eqnarray} 
 From this equation, the expansion of $k^2$ 
in terms of the free particle solution $\kappa$ 
define by 
\begin{eqnarray} 
\label{4.5} 
-{{d^2}\over{dx^2}} \varphi (x) 
= \kappa^2 \varphi (x) 
\end{eqnarray} 
is easily obtained as 
\begin{eqnarray} 
\label{4.6} 
k^2 \approx \kappa^2 + {2 \over L} v 
           - {1\over{L^2 \kappa^2}} v^2 + \cdots . 
\end{eqnarray} 
Similarly, for $V(x) = c_a E_a(x)$ at $a \to 0$, 
the connection condition Eq. (\ref{2.17}) gives 
\begin{eqnarray} 
\label{4.7} 
\tan{ {{k L}\over 2} } = - {k c\over 2} ,
\end{eqnarray} 
which yields 
\begin{eqnarray} 
\label{4.8} 
k^2 \approx \kappa^2 - {{2\kappa^2} \over L} c 
           + {{3\kappa^2}\over  L^2} c^2 + \cdots . 
\end{eqnarray} 
Suppose we are interested in obtaining these results 
through perturbation theory that starts from 
the free problem, Eq. (\ref{4.5}) 
whose solution is given by 
\begin{eqnarray} 
\label{4.9} 
\varphi_n(x) &=& {1\over\sqrt{L}}{\rm e}^{i\kappa_n x}, 
\\ \nonumber 
\kappa_n &=& {{2\pi}\over L} n 
\ \ \ \ \ \ \ \ (n=0,\pm1,\pm2,\cdots). 
\end{eqnarray} 
The Rayleigh-Schr{\"o}dinger formula reads 
\begin{eqnarray} 
\label{4.10} 
k_n^2 \approx \kappa_n^2 
+ \left\langle n \right| V 
  \left| n \right\rangle 
+ \sum_m 
 {' 
  { 
    {\left\langle n \right| V 
     \left| m \right\rangle 
     \left\langle m \right| V 
     \left| n \right\rangle} 
   \over 
    {\kappa_n^2-\kappa_m^2} 
  } 
 } 
\cdots , 
\end{eqnarray} 
where $\sum{'}$ represents the exclusion of $m=n$, and 
the matrix elements are given by 
\begin{eqnarray} 
\label{4.11} 
\left\langle m \right| V \left| n \right\rangle 
\equiv 
\int_{-\infty}^{\infty}{dx\varphi_m^*(x)V(x)\varphi_n(x)} . 
\end{eqnarray} 
Because of the degeneracy $\kappa_{-n} = \kappa_n$, 
it is advantageous to use the symmetrized and antisymmetrized 
wave functions. 
\begin{eqnarray} 
\label{4.12} 
\varphi_{n\pm} 
\equiv {1 \over \sqrt{1+\delta_{n0}}}\cdot {1 \over \sqrt{2}} 
       \left( \varphi_n \pm \varphi_{-n} \right) . 
\end{eqnarray} 
Because of the mirror symmetry $V(-x) = V(x)$, one has 
\begin{eqnarray} 
\label{4.13} 
\left\langle m_\pm \right| V \left| n_\mp \right\rangle 
= 0 
\end{eqnarray} 
which guarantees the separation of the state vectors into the 
$\{ n_+ \}$ and $\{ n_- \}$ sectors.  within each sector, one has 
\begin{eqnarray} 
\label{4.14} 
\left\langle m_\pm \right| V \left| n_\pm \right\rangle 
=  {1 \over \sqrt{(1+\delta_{n0})(1+\delta_{m0})}} 
\\ \nonumber 
\times 
\left\{ 
\left\langle m \right| V \left| n \right\rangle 
\pm \left\langle m \right| V \left| -n \right\rangle 
\right\} . 
\end{eqnarray} 
Explicit forms of the matrix elements, Eq. (\ref{4.11}) are given by 
\begin{eqnarray} 
\label{4.15} 
\left\langle m_\pm \right| \Delta_a \left| n_\pm \right\rangle 
&=& {1 \over \sqrt{(1+\delta_{n0})(1+\delta_{m0})}} \cdot 
    {1\over L} 
\\ \nonumber 
& &\times 
\left( 
      {{\sin{a\kappa^-_{nm}}}\over{a\kappa^-_{nm}}}\pm 
      {{\sin{a\kappa^+_{nm}}}\over{a\kappa^+_{nm}}} 
    \right) , 
\end{eqnarray} 
\begin{eqnarray} 
\label{4.16} 
\left\langle m_\pm \right| E_a \left| n_\pm \right\rangle 
&=& {-1 \over \sqrt{(1+\delta_{n0})(1+\delta_{m0})}} \cdot 
    {\kappa_m\kappa_n\over L} 
\\ \nonumber 
& &\times 
    \left( 
        {{\sin{a\kappa^-_{nm}}}\over{a\kappa^-_{nm}}}\mp 
        {{\sin{a\kappa^+_{nm}}}\over{a\kappa^+_{nm}}} 
    \right) , 
\end{eqnarray} 
where the notation 
\begin{eqnarray} 
\label{4.17} 
\kappa^{\pm}_{nm} \equiv \kappa_n \pm \kappa_m 
\end{eqnarray} 
is adopted, and it is understood that for $m=n$, 
$(\sin{a\kappa^-_{mn}})/(a\kappa^-_{mn})$ is set to be $1$. 
 From these expressions, 
it is easy to see that at $a \to 0$ limit, we have 
\begin{eqnarray} 
\label{4.18} 
\left\langle m_- \right| \Delta_a \left| n_- \right\rangle 
\rightarrow 0 , 
\\ 
\left\langle m_+ \right| E_a \left| n_+ \right\rangle 
\rightarrow 0 . 
\end{eqnarray} 
Therefore, it is sufficient to consider only $\{n_+\}$ for $\Delta_a$ 
and $\{n_-\}$ for $E_a$. 
Note the fact that for $\{n_-\}$, only $n \ge 1$ is allowed, 
while for $\{n_+\}$, $n$ can be any non-negative value 
including 0. 

We start with the case of $\Delta_a$. 
Assuming the unperturbed state is of positive energy $\kappa_n > 0$ 
(therefore $n\ne 0$), one has 
\begin{eqnarray} 
\label{4.20} 
\left\langle n_+ \right| \Delta_a \left| n_+ \right\rangle 
= {2\over L} , 
\end{eqnarray} 
and 
\begin{eqnarray} 
\label{4.21} 
& & 
\sum_m{' 
  { 
    {| \left\langle n_+ \right| \Delta_a 
       \left| m_+ \right\rangle | ^2} 
   \over 
    {\kappa_n^2-\kappa_m^2} 
  }   } 
\\ \nonumber 
&=& {1\over L^2} 
    \sum_{m=0}^{\infty}{' 
       {1\over{\kappa_n^2-\kappa_m^2}} 
       {\left( 
         {{\sin{a\kappa^-_{nm}}}\over{a\kappa^-_{nm}}}+ 
         {{\sin{a\kappa^+_{nm}}}\over{a\kappa^+_{nm}}} 
        \right)^2} 
       {1\over{1+\delta_{m0}}} 
                       } 
\\ \nonumber 
&\approx& {1\over L^2} 
    \left[ 
    \sum_{m=1}^{\infty}{' 
       {4\over{\kappa_n^2-\kappa_m^2}} 
                       } + {2\over\kappa_n^2} 
    \right] 
\\ \nonumber 
&=& -{1\over{L^2\kappa_n^2}} ,
\end{eqnarray} 
where the limit $a \to 0$ is taken at the second line, and 
a well known summation formula is used in the last line. 
Thus one recovers the correct expansion, Eq. (\ref{4.5}), 
which is of course a well-known result 
from elementary textbooks. 

Now we consider the perturbation by $E_a$.  The first order term 
is easily obtained as 
\begin{eqnarray} 
\label{4.22} 
\left\langle n_- \right| E_a \left| n_- \right\rangle 
= -{2\kappa_n^2\over L} , 
\end{eqnarray} 
The second order term is given by 
\begin{eqnarray} 
\label{4.23} 
& & 
\sum_m{' 
  { 
    {|\left\langle n_- \right| E_a 
      \left| m_- \right\rangle | ^2} 
   \over 
    {\kappa_n^2-\kappa_m^2} 
  }   } 
\\ \nonumber 
&=& {1\over L^2} 
    \sum_{m=1}^{\infty}{' 
       {{\kappa_n^2\kappa_m^2}\over{\kappa_n^2-\kappa_m^2}} 
       {\left( 
         {{\sin{a\kappa^-_{nm}}}\over{a\kappa^-_{nm}}}+ 
         {{\sin{a\kappa^+_{nm}}}\over{a\kappa^+_{nm}}} 
        \right)^2} 
                       } 
\\ \nonumber 
&=& {1\over L^2} 
    \left[ 
    \kappa_n^4 
    \sum_{m=1}^{\infty}{' 
       {1\over{\kappa_n^2-\kappa_m^2}} 
       {\left( 
         {{\sin{a\kappa^-_{nm}}}\over{a\kappa^-_{nm}}}+ 
         {{\sin{a\kappa^+_{nm}}}\over{a\kappa^+_{nm}}} 
        \right)^2} 
                       } 
    \right. 
\\ \nonumber 
& & \ \ \ 
    \left. 
   -\kappa_n^2 
    \sum_{m=1}^{\infty}{' 
       {\left( 
         {{\sin{a\kappa^-_{nm}}}\over{a\kappa^-_{nm}}}+ 
         {{\sin{a\kappa^+_{nm}}}\over{a\kappa^+_{nm}}} 
        \right)^2} 
                       } 
    \right] . 
\end{eqnarray} 
In the second equation, the first term can be 
readily calculated by taking the $a \to 0$ limit 
as in the case of $\Delta_a$: 
\begin{eqnarray} 
\label{4.24} 
& & 
    \sum_{m=1}^{\infty}{' 
       {1\over{\kappa_n^2-\kappa_m^2}} 
       {\left( 
         {{\sin{a\kappa^-_{nm}}}\over{a\kappa^-_{nm}}}+ 
         {{\sin{a\kappa^+_{nm}}}\over{a\kappa^+_{nm}}} 
        \right)^2} 
                       } 
\\ \nonumber 
&\approx& 
    \sum_{m=1}^{\infty}{' 
       {4\over{\kappa_n^2-\kappa_m^2}} 
                       } 
\\ \nonumber 
&=& -{3\over \kappa_n^2} . 
\end{eqnarray} 
The divergence of the second term at $a \to 0$ limit has to 
be explicitly handled. Using the relation in the Appendix, 
one obtains 
\begin{eqnarray} 
\label{4.25} 
\sum_{m=1}^{\infty}{' 
     {\left( 
       {{\sin{a\kappa^-_{nm}}}\over{a\kappa^-_{nm}}}+ 
       {{\sin{a\kappa^+_{nm}}}\over{a\kappa^+_{nm}}} 
      \right)^2} 
                   } 
&\approx& {{2\pi}\over{2\pi a/L}} - 6 
\\ \nonumber 
&=& {L\over a} - 6 . 
\end{eqnarray} 
We have 
\begin{eqnarray} 
\label{4.26} 
\sum_m{' 
  { 
    {|\left\langle n_- \right| E_a 
      \left| m_- \right\rangle | ^2} 
   \over 
    {\kappa_n^2-\kappa_m^2} 
  }   } 
\approx 3{\kappa_n^2 \over L^2} - {{\kappa_n^2 L}\over a} . 
\end{eqnarray} 
Therefore, the summation to the second order, Eq. (\ref{4.10}) gives 
\begin{eqnarray} 
\label{4.27} 
k^2 \approx \kappa^2 
            \{ 1-{2\over L}c_a+{3\over L^2}c_a^2-{1\over{L a}}c_a^2 
            \} . 
\end{eqnarray} 
The appearance of the divergence calls for the renormalization 
procedure.  The replacement of bare coupling $c_a$ by 
the renormalized coupling 
$c$ serves the purpose. 
Using the expansion of Eq. (\ref{2.18}) in terms of $1/a$ 
\begin{eqnarray} 
\label{4.28} 
c_a = {{2ac} \over{2a +c}} \approx c -{c^2\over{2a}} + \cdots , 
\end{eqnarray} 
we obtain 
\begin{eqnarray} 
\label{4.29} 
k^2 \approx \kappa^2 
            \{ 1-{2\over L}c+{1\over{L a}}c^2 
                +{3\over L^2}c^2-{1\over{L a}}c^2 
            \} . 
\end{eqnarray} 
Thus the divergence is cancelled out, 
and the correct result Eq. (\ref{4.7}) is reproduced. 

It should be possible 
to go on to higher order of perturbation 
and to show the cancellation of divergent term order by order. 
Also, perturbative calculation of wave function could be 
similarly performed.  But the calculation up to the second order 
shown here would be sufficient to convince the validity 
of perturbative renormalization in this model. 

An interesting fact of the perturbative expansion for $c_a E_a(x)$, 
Eq. (\ref{4.8})  is 
that it is formally identical to the strong coupling expansion 
of $v \Delta_a(x)$ in terms of $1/v$. 
This, of course, is not an accident 
but the result of the duality between $\delta$ and $\varepsilon$ 
potentials \cite{CS98a,CS98b}. 

\bigskip 
\section{Conclusion} 

Now that a calculational scheme is devised to deal with the 
divergences arising in the perturbative calculation of 
discontinuity-inducing interaction, we stand at the starting 
line to tackle more practical physical problems. 
Simplest among them is the system 
that has both regular finite-range potential 
{\it and} the contact force.  More interesting is the finite 
range potential problem that inherently requires 
the consideration of 
the singular zero-range force, that of $1/|x|$, or 
``one-dimensional Coulomb'' problem as is sometimes called 
\cite{LO59,GC97}. 
Our approach would be profitably applied also to the 
many-body problem of one-dimensional particles where the 
second-quantized representation is frequently utilized. 
Another direction for the potential development is the 
generalization of our analysis to the relativistic quantum 
mechanical models, and ultimately to the field theoretical models. 
Such analysis might be useful to shed some light on the 
relationship between the fermion-boson 
dualities found in quantum mechanics and 
in field theories \cite{CS98a,CO75}. 
It is known that one can construct three parameter family of 
generalized contact interaction by combining the $\delta$ 
and $\varepsilon$ interactions \cite{CS98}. 
A detailed study of most general contact interaction in one 
dimensional quantum mechanics and its possible relativistic 
and field theoretical extension should be also of great interest. 

Prior to the proper formulation of the discontinuity-inducing 
contact force, there has been an introduction of a set of 
interactions known in nuclear physics as 
Skyrme force \cite{SK56} 
which includes a component analogous to prime-delta-prime 
force, Eq. (\ref{2.20}) in {\it spatial dimension three}. 
With such interactions, 
a rather detailed numerical analysis has been carried 
out \cite{NV72}.  However, 
from the current view, 
it is clear that 
the results drawn from such analysis are in need of 
critical reexamination, since 
such object cannot be defined as renormarizable interaction 
in other dimensions than one.  This is to be contrasted 
to the $\delta$ interaction which {\em can} be 
constructed as renormalizable interactions in 
dimensions two and three \cite{AG88}. 

Throughout our treatment, we have had 
no need to invoke any of the high mathematics traditionally 
associated with the theory of generalized contact interactions. 
In fact, it has been felt that the current analysis might 
be useful as pedagogical materials for 
the graduate level textbook on quantum mechanics. 
Specifically, the great simplification garnered by 
the solvability of problems helps detach 
the concept of the renormalization from 
the complex machinery of field theory. 
It enables one to readily recognize that the two facets 
of renormalization, namely, 
the invariance of the problem with 
suitable rescaling, and a technical procedure 
that allows the perturbative treatment, 
are one and the same thing. 

Finally we would like to reiterate our basic message 
that 
the discontinuity-inducing interaction, a lesser known twin 
of $\delta$ function, now calls for a proper recognition as 
an integral element of the quantum mechanics. 

\bigskip 
\acknowledgements 
%
This work has been supported in part by the Grant-in-Aid 
(No. 10640396) by the Japanese Ministry of Education. 

\bigskip 
\bigskip 
\bigskip 
%
\appendix{\small\bf 
\ \ \ \ \ \ \ \ 
\ \ \ \ \ \ \ \ 
\ \ \ \ \ \ \ 
APPENDIX} 

\bigskip 
We outline the proof of the summation formula 
\begin{eqnarray} 
\label{B1} 
& & 
{1\over\beta^2} 
\sum_{m=1}^{\infty}{' 
     {\left( 
       {\sin{\beta(n-m)}\over{n-m}}+ 
       {\sin{\beta(n+m)}\over{n+m}} 
      \right)^2} 
                   } 
\\ \nonumber 
&=& {{2\pi}\over{\beta}} - 6 + O(\beta) 
\end{eqnarray} 
in the limit $\beta \to 0$. 
It can be split into two steps: 
First, we have a well known relation 
\begin{eqnarray} 
\label{B2} 
\sum_{m=1}^{\infty} 
       {\sin^2{\beta m}\over{m^2}} 
= {1\over 2}\beta(\pi-\beta) . 
\end{eqnarray} 
Next we can show, for small $\beta$, 
\begin{eqnarray} 
\label{B3} 
& & 
{1\over\beta^2} 
\sum_{m=1}^{\infty}{' 
     {\left( 
       {\sin{\beta(n-m)}\over{n-m}}+ 
       {\sin{\beta(n+m)}\over{n+m}} 
      \right)^2} 
                   } 
\\ \nonumber 
& & \ \ \ \ \ \ 
+4 
- 
{4\over\beta^2} 
\sum_{m=1}^{\infty} 
       {\sin^2{\beta m}\over{m^2}} 
\\ \nonumber 
&\approx& {1\over\beta} 
  \int_0^\infty{dx 
                \left[ 
                  \left( 
                      {\sin{(\beta n-x)}\over{\beta n-x}} 
                     +{\sin{(\beta n+x)}\over{\beta n+x}} 
                  \right)^2 
                \right.} 
\\ \nonumber 
& & \ \ \ \ \ \ \ \ \ \ \ \ \ \ \ \ 
               {\left. 
                  -{4\sin^2{x}\over x^2} 
                \right] 
               } 
\\ \nonumber 
& &\ \ \ \ 
 +\left[ 
     4- \left( 1+{\sin{2\beta n}\over{2\beta n}} \right)^2 
  \right] 
\\ \nonumber 
&=& 
O(\beta) . 
\end{eqnarray} 
It is easy to see that one obtains Eq. (\ref{B1}) 
by combining these two equations. 
%

 
%
%
\end{document}